\documentstyle[12pt,epsfig]{article}

\begin{document}
\begin {center}
{\bf {\Large
Absorption of $\omega$ and $\phi$ mesons in the inclusive photonuclear
reaction
} }
\end {center}
\begin {center}
Swapan Das \footnote {email: swapand@barc.gov.in} \\
{\it Nuclear Physics Division,
Bhabha Atomic Research Centre,  \\
Trombay, Mumbai: 400085, India \\
Homi Bhabha National Institute, Anushakti Nagar,
Mumbai-400094, India }
\end {center}

\begin {abstract}
The $\omega$ and $\phi$ mesons dominantly decay outside the nucleus because
the decay-lengths of these mesons are much larger than the dimension of the
nucleus. Therefore, the medium modification of both $\omega$ and $\phi$
mesons is studied through their absorption in the nucleus. 
To
look for the medium modification of the $\omega$ and $\phi$ mesons, CLAS
collaboration at Jeffereson Laboratory (Jlab) has investigated the absorption
of these mesons by measuring their nuclear transparency ratios in the
$\gamma A \to \omega (\phi) X \to e^+e^-X$ reaction.
The
quoted transparency ratios have been calculated, and those are compared with
the experimental findings reported from Jlab.
\end {abstract}


PACS number(s): 13.60.Le, 25.20.-x

\section{Introduction}

The modification of the in-medium properties of vector meson, e.g., $\rho$,
$\omega$, $\phi$, ... mesons, is related to the chiral symmetry restoration
in the nuclear medium \cite{hats}.
Large
medium effect on the vector meson is expected in the heavy-ion collision
which occurs far from equilibrium, over a wide range of temperature and
density. Therefore, it is difficult to interpret the relation between the
chiral symmetry restoration and modification of the vector meson in the
heavy-ion collision \cite{agak} (see the references there).
The
change in the hadronic properties of vector meson in the nucleus (normal
and cold) is predicted significant \cite{effen} and, also the chiral
symmetry is shown to restore partially in the nucleus \cite{birse}.
Since
the modification of the vector meson in the nucleus involves less complexity
compared to that in the heavy-ion collision, the in-medium properties of
the vector meson in the nucleus can be studied conclusively.

The search for the vector meson modification in the nucleus, both
theoretically and experimentally, is carried out extensively for last few
decades.
Apart from those mentioned above \cite{effen}, there exist calculations
which report the medium effect on the vector meson in the nucleus.
The
modification of the $\rho$ meson in the pion nucleus reaction is described
in Ref.~\cite{golu}. The in-medium properties of $\rho$ and $\omega$ mesons
in the $\gamma A$ reaction are reported \cite{das, mueh, das2}, and those of
the $\phi$ meson produced in the nuclear reactions is also studied
\cite{cabr, gubl}.
CBELSA/TAPS
collaboration looked for the modification of $\omega$ meson in the
photonuclear reaction \cite{trnka}. KEK-PS E325 collaboration reported the
reduction of the vector meson mass in $pA$ reaction \cite{naruki}. The
measured nuclear transparencies emphasized large in-medium width of the
$\omega$ and $\phi$ mesons \cite{kotu, poly}.
A
recent review on the meson nucleus interactions and in-medium modification
of mesons can be seen in Ref.~\cite{metag}.

The precise measurement was done by CLAS collaboration at Jlab \cite{nasse,
wood} to search the in-medium properties of vector meson (i.e., $\rho^0$,
$\omega$ and $\phi$ mesons) in the inclusive $A(\gamma, e^+e^-)X$ reaction,
where $X$ denotes the undetected final state of the nucleus.
In
this measurement, the Bremsstrahlung photon beam of energy up to 4 GeV was
used to produce the vector mesons, and those were detected by their
decay product $e^+e^-$ in the final state.
The
momenta of the vector mesons were restricted greater than 0.8 GeV/c. The
use of electromagnetic probe in this experiment provides undistorted
information about the hadronic properties of vector meson in the nucleus.
Being
a short-lived ($\tau \sim 10^{-23}$ sec) particle, the $\rho$ meson
dominantly decays inside the nucleus. Broadening and no mass-shift of this
meson has been observed in the experiment \cite{nasse}. The calculated
results also corroborate this finding \cite{das}.

The $\omega$ and $\phi$ mesons, unlike the $\rho$ meson, dominantly decay
outside the nucleus \cite{das2}, as they possess large decay-lengths (i.e.,
$\sim 23$ fm and $\sim 44$ fm respectively) compared to the dimension of a
nucleus. Therefore, the medium effect on these mesons has been studied at
Jlab \cite{wood} by measuring the nuclear transparency (which describes the
absorption) of these mesons.
The
experimental results show that the absorption of the $\omega$ meson in
nuclei cannot be explained by the theoretical models considering the drastic
increase in the $\omega$ meson nucleon cross section in the nucleus.
Contrary
to this, the absorption of $\phi$ meson in nuclei has been understood by the
large in-medium $\phi$ meson nucleon cross section, i.e.,
$\sigma_t^{*\phi N}=16-70$ mb.
In
the nucleus, the enhancement in the vector meson nucleon cross section
$\sigma_t^{*VN}$ modifies the collision width $\Gamma_{V,c}$ of the vector
meson: $\Gamma_{V,c} \propto \sigma_t^{*VN}$.
It
should be mentioned that the (semi)hadronic decay products of the $\omega$
and $\phi$ mesons are not absorped in the nucleus, as these mesons (as
mentioned above) dominantly decay outside the nucleus. Therefore, the
(semi)hadronic decay channels are also partinent probes for studyng the
nuclear transparency of the $\omega$ and $\phi$ mesons \cite{kotu, ishi}.

To investigate the issues of $\omega$ and $\phi$ mesons' absorption in the
nucleus, the transparency of these mesons in the inclusive $(\gamma, V)$
reaction on nuclei is studied. $V$ (meson) represents a vector meson, e.g.,
$\omega$ or $\phi$ meson.
Since
the final state of the nucleus is not specified in the inclusive reaction,
the sum over all nuclear final states is taken to calculate the cross section
of the reaction.
The
absorption of these mesons arises due to their interaction with the nucleus,
which is described by the $V$ meson nucleus optical potential $V_{OV}$.
The energy dependent $V$ meson nucleon cross section is used to evaluate
$V_{OV}$.
The
distributed mass of $V$ meson is illustrated by its mass distribution
function. Since this meson is detected by the dielectron decay
(i.e., $V \to e^+e^-$), the mass dependent decay of it is estimated
by the  branching ratio $\Gamma_{V \to e^+e^-} (m) / \Gamma_V (m)$.

\section{Formalism}

The cross section of the inclusive $(\gamma,V)$ reaction on a nucleus,
i.e., $A(\gamma,V)X$ reaction, can be written as
\begin{equation}
\sigma_t^{\gamma A \to VX} (m,E_\gamma)
=\int d\Omega_V K_F <|{\cal M}_{fi}|^2>,
\label{gAX}
\end{equation}
where $m$ is the mass of vector meson and $E_\gamma$ is the beam energy.
$K_F$ arises due to phase space of the reaction:
$K_F =\frac{\pi^2}{(2\pi)^4}
\frac{k^2_V E_X}{k_\gamma |k_VE_i-k_\gamma E_Vcos\theta_V|}$. $\Omega_V$ is
the solid angle associated with the vector meson momentum.

The matrix element of the considered reaction, i.e., ${\cal M}_{fi}$, is
given by
\begin{equation}
{\cal M}_{fi} = \sum_i
<X; \chi^{(-)}({\bf k}_V, {\bf r})
    \epsilon_{V,\mu} ( {\bf k}_V, {\bf \lambda}_V)
 | \tilde{f}^i_{\gamma N \to VN} |
   \epsilon^\mu_\gamma ({\bf k_\gamma,\lambda_\gamma})
    e^{i{\bf k}_\gamma . {\bf r}}; A>,
\label{mfi}
\end{equation}
where $|A>$ and $|X>$ are the initial and final (undetected) states of the
nucleus. $\chi^{(-)}({\bf k}_V, {\bf r})$ represents the distorted wave
function of the vector meson. $\epsilon_{\gamma (V)}$ is the polarization
vector of $\gamma (V)$, and $\lambda_{\gamma(V)}$ denotes the polarization
component.
$\sum_i$ denotes the summation over all nucleons in the nucleus.
$\tilde{f}^i_{\gamma N \to VN}$ is given by
$\tilde{f}^i_{\gamma N \to VN} =-4\pi  ( 1 + \frac{E_\gamma}{E_N} )
f^i_{\gamma N \to VN} (0)$  \cite{frank}, where
$f^i_{\gamma N \to VN} (0)$ denotes the forward amplitude of the
$\gamma N \to VN$ reaction.

The annular bracket around $|{\cal M}_{fi}|^2$ in Eq.~(\ref{gAX}) represents
the average over the spins in the initial state and the sum over the final
states. In fact, $|{\cal M}_{fi}|^2$ has to be multiplied by the phase-space
and incident flux factors of the reaction, then the summation over the
nuclear final state $X$ has to be taken to get the cross section of the
inclusive reaction.
Since
the energy of the outgoing vector meson is in the GeV region, the energy of
this meson can be considered essentially independent of $X$. It follows the
phase-space factor is independent of $X$ \cite{howell}. Therefore,
$<|{\cal M}_{fi}|^2>$ can be written as
\begin{eqnarray}
<|{\cal M}_{fi}|^2>
&=& \frac{1}{2} \sum_{\lambda_\gamma \lambda_V}
    \sum_X |{\cal M}_{fi}|^2,  \nonumber  \\
&=& \frac{1}{2} \sum_{\lambda_\gamma \lambda_V}
    |\epsilon^*_{V,\mu} ( {\bf k}_V, {\bf \lambda}_V) ~
     \epsilon^\mu_\gamma ({\bf k_\gamma,\lambda_\gamma}) |^2
     \sum_X |F_{XA}|^2.
\label{mfi2}
\end{eqnarray}

$F_{XA}$ describes the matrix element:
\begin{equation}
F_{XA} = \sum_i   <X; \chi^{(-)}({\bf k}_V, {\bf r}) |
        \tilde{f}^i_{\gamma N \to VN} | e^{i{\bf k}_\gamma . {\bf r}}; A>.
\label{fn0}
\end{equation}
The distorted wave function of the vector meson
$\chi^{(-)}({\bf k}_V, {\bf r})$, according to the eikonal description
in Glauber model \cite{glaub}, is given by
\begin{equation}
\chi^{(-)}({\bf k}_V, {\bf r}) = e^{i {\bf k}_V . {\bf r}}
exp \left [ \frac{i}{v}
\int_z^\infty  V^*_{OV} ({\bf b},z^\prime) dz^\prime \right ].
\label{dwv}
\end{equation}
$V_{OV}$ denotes the vector meson nucleus optical potential
\cite{glaub}:
\begin{equation}
\frac{1}{v}V_{OV} ({\bf r^\prime})
= -\frac{1}{2} (\alpha_{VN}+i) \sigma_t^{VN} \varrho ({\bf r^\prime}),
\label{IVop}
\end{equation}
where $\varrho ({\bf r^\prime})$ represents the density distribution of the
nucleus, normalized to the mass number of the nucleus. $\alpha_{VN}$ is the
ratio of the real to imaginary part of the elementary $V$ meson nucleon
$(VN)$ scattering amplitude, and $\sigma_t^{VN}$ is the elementary
$VN$ total scattering cross section.

Considering the independent particle model of the nucleus \cite{bauer} and
$f^i_{\gamma N \to VN}$ identically same for all nucleons in the nucleus,
$\sum_X |F_{XA}|^2$ can be written as
\begin{equation}
\sum_X |F_{XA}|^2 = |\tilde{f}^i_{\gamma N \to VN}|^2 \int d{\bf r} ~
exp \left [ \frac{2}{v}
\int^\infty_z W_{OV} ({\bf b},z^\prime) dz^\prime \right ] \varrho ({\bf r}),
\label{fxA}
\end{equation}
where $W_{OV}$ denotes the imaginary part of $V_{OV}$.

The vector meson produced in the $(\gamma, V)$ reaction possesses
distributed mass $m$ and it is detected by the dielectron decay, i.e.,
$V \to e^+e^-$. Therefore, the cross section of the considered reaction
(i.e., $\gamma A \to VX$; $V \to e^+e^-$) can be written as
\begin{equation}
\sigma_t^{\gamma A \to e^+e^-X} (E_\gamma)
=\int^{m_{mx}}_{m_{mn}} dm^2
 \frac{ \Gamma_{V\to e^+e^-}(m) }{ \Gamma_V(m) } S(m)
 \sigma_t^{\gamma A \to VX} (m, E_\gamma),
\label{gAX2}
\end{equation}
where
$\Gamma_V (m)$ is the total decay width of the vector meson.
$\Gamma_{V\to e^+e^-} (m)$ illustrates the width for the decay:
$V\to e^+e^-$. The cross section $\sigma_t^{\gamma A \to VX} (m, E_\gamma)$
is already mentioned, see Eq.~(\ref{gAX}).
$S(m)$
represents the mass distribution function of the vector meson, which is
described by the Breit-Wigner form \cite{brat}:
\begin{equation}
S(m)= \frac{1}{\pi}
      \frac{ m_V\Gamma_V(m) }{ (m^2-m^2_V)^2 + m^2_V \Gamma^2_V(m) },
\label{mdf}
\end{equation}
where $m_V$ is the pole mass of the vector meson.

For Bremsstrahlung photon induced reaction, the beam possesses a certain
energy range and the Bremsstrahlung cross section varies as
$\frac{1}{E_\gamma}$ \cite{sober}. Therefore, the cross section of the
Bremsstrahlung beam induced $\gamma A \to e^+e^-X$ reaction
can be written \cite{kask} as
\begin{equation}
\sigma_t^{\gamma A \to e^+e^-X}
= \int^{E_{mx}}_{E_{mn}} dE_\gamma  W(E_\gamma)
  \sigma_t^{\gamma A \to e^+e^-X} (E_\gamma);
\label{gAX5}
\end{equation}
with $W(E_\gamma) \propto \frac{1}{E_\gamma}$. The beam energy
dependent cross section $\sigma_t^{\gamma A \to e^+e^-X} (E_\gamma)$ is
given in Eq.~(\ref{gAX2}).

The nuclear transparency $T_A$ of the vector meson in the
$\gamma A \to e^+e^-X$ reaction has been calculated to study the absorption
of vector meson in nuclei. It is defined \cite{das, cabr} as 
\begin{equation}
T_A \simeq \frac{ \sigma_t^{\gamma A \to e^+e^-X} }
          { \sigma_{t, vac}^{\gamma A \to e^+e^-X} }.
\label{ta}
\end{equation}
$\sigma_t^{\gamma A \to e^+e^-X}$ is described in Eq.~(\ref{gAX5}), and
$\sigma_{t, vac}^{\gamma A \to e^+e^-X}$ represents that where the vector
meson optical potential, i.e., $V_{OV}$ in Eq.~(\ref{dwv}), is taken equal
to zero.

\section{Result and Discussions}

The vector meson production amplitude in the $\gamma N \to VN$ reaction,
i.e., $f_{\gamma N \to VN}$, is related to the four-momentum transfer
distribution $\frac{d\sigma}{dq^2}$ of that reaction \cite{sibir} as
\begin{equation}
\frac{d\sigma}{dq^2} (\gamma N \to VN)
= \frac{\pi}{k^2_\gamma} |f_{\gamma N \to VN}|^2.
\label{fmt}
\end{equation}
Using this equation, the magnitude of the forward amplitude, i.e.,
$f_{\gamma N \to VN} (0)$, is extracted from the measured forward
$\frac{d\sigma}{dq^2}$ for both $\omega$ and $\phi$ mesons
\cite{sibir, barth}, and that has been used to calculate the cross section
of the $\gamma A \to VX$ reaction.

The absorption of the vector meson in the nucleus arises due to the
imaginary part of the vector meson nucleus optical potential
$W_{OV} ({\bf r})$ in Eq.~(\ref{fxA}), which is described by the vector
meson nucleon total scattering cross section $\sigma_t^{V N}$, see
Eq.~(\ref{IVop}).
The
$\omega N$ total scattering cross section $\sigma_t^{\omega N}$, given in
Ref.~\cite{sibir02}, is used to evaluate the imaginary part of the $\omega$
meson optical potential $W_{O\omega}$.
For
$\phi$ meson, $\frac{d\sigma}{dq^2} (\gamma N \to \phi N) |_{q^2=0}$
(according to vector meson dominance model \cite{sibir, das6})
is given by
\begin{equation}
\frac{d\sigma}{dq^2} (\gamma N \to \phi N) |_{q^2=0}
= \frac{\alpha_{em}}{16\gamma^2_{\gamma \phi}}
  \left ( \frac{ \tilde{k}_\phi }{ \tilde{k}_\gamma } \right )^2
   [1+\alpha^2_{\phi N}] (\sigma^{\phi N}_t)^2,
\label{ftD2}
\end{equation}
where $\alpha_{em} (=1/137.036)$ is the fine structure constant.
$\gamma_{\gamma \phi}$ is the photon $\gamma$ to $\phi$ meson coupling
constant which is determined from the measured $\phi \to e^+e^-$
decay width \cite{olive}:  $ \gamma_{\gamma \phi} = 6.72 $ \cite{das6}.
$\tilde{k}_\phi$
and $\tilde{k}_\gamma$ are the momenta in the $\phi N$ and $\gamma N$ c.m.
systems respectively, evaluated at the c.m. energy of $\gamma N$ system.
$\alpha_{\phi N}$
denotes the ratio of the real to imaginary part of the $\phi N$ scattering
amplitude, and $\sigma^{\phi N}_t$ is the total $\phi N$ scattering cross
section.
Using
the measured $\frac{d\sigma}{dq^2} (\gamma N \to \phi N) |_{q^2=0}$
\cite{sibir} and $\alpha_{\phi N} =-0.3$ \cite{bauer}, the magnitude of
$\sigma^{\phi N}_t$ is determined. The momentum dependent values of
$\sigma^{\omega N}_t$ (for $m_\omega =782.65$ MeV) and $\sigma^{\phi N}_t$
(for $m_\phi =1.02$ GeV) are given in Table~1.

\begin{table}
\begin{center}
\caption {
Table~1: The momentum dependent vector meson nucleon total cross sections.
$\sigma_t^{\omega N}$ is based on Ref.~\cite{sibir02}. 
$\sigma_t^{\phi N}$ is described in text.
}
\begin{tabular} { c| c| c}
\hline
$k_{\omega (\phi)}$ & $\sigma_t^{\omega N}$ & $\sigma_t^{\phi N}$   \\  
(GeV/c)             &  (mb)                 &   (mb)    \\  \hline 
1.0                 &  79.18                &   13.07   \\ 
1.5                 &  64.78                &   11.79   \\ 
2.0                 &  56.18                &   11.25   \\ 
2.5                 &  50.72                &   11.00   \\ 
3.0                 &  46.80                &   10.82   \\ 
3.5                 &  43.99                &   10.53   \\ 
4.0                 &  41.89                &   ~10.35   
\label{TbXT}
\end{tabular}
\end{center}
\end{table}

The density distribution of the nucleus $\varrho ({\bf r})$ of deuteron is
evaluated using its wave function due to paris potential \cite{lacom}. The
distribution of $\varrho ({\bf r})$ for other nuclei, as extracted from the
electron scattering data, is given in Ref.~\cite{andt}.
$\varrho (r)$
for $^{12}$C nucleus is described by the modified harmonic-oscillator form,
whereas that for $^{56}$Fe, $^{112}$Sn and $^{208}$Pb nuclei is illustrated
by the two-parameter Fermi distribution function. The parameters appearing
in $\varrho (r)$ are duly quoted in Ref.~\cite{andt}.

The width of the vector meson $\Gamma_V (m)$ is composed of the partial
decay widths \cite{olive}, i.e.,
$\Gamma_\omega (m)=$
$  \Gamma_{\omega \to \pi^+\pi^-\pi^0} (m)
+  \Gamma_{\omega \to \pi \gamma} (m)
+  \Gamma_{\omega \to \pi^+\pi^-} (m)
+  \Gamma_{\omega \to e^+e^-} (m) $,
and
$\Gamma_\phi (m) =$
$  \Gamma_{\phi \to K^+K^-} (m)
+  \Gamma_{\phi \to {\bar K}^0K^0} (m)
+  \Gamma_{\phi \to \rho \pi} (m)
+  \Gamma_{\phi \to e^+e^-} (m) $.
The
width $\Gamma_{\omega \to \pi^+\pi^-\pi^0} (m)$, as described by Sakurai, is
given in Ref.~\cite{saku}. The twobody hadronic and semihadronic decay
width of the vector meson, e.g., $V \to h_1h_2$, can be expressed
\cite{manl} as
\begin{equation}
\Gamma_{V \to h_1h_2} (m)
= \Gamma_{V \to h_1h_2} (m_V) \frac{\Phi_l(m)}{\Phi_l(m_V)};
~~~ \Phi_l(m) = \frac{\tilde {k}}{m} B^2_l (\tilde{k}R),
\label{wvm}
\end{equation}
with $R=1$ fm \cite{pary0}. $\Gamma_{V \to h_1h_2} (m_V)$ is duly tabulated
in Ref.~\cite{olive}. $\tilde {k}$ is the momentum of $h_1$ (or $h_2$)
originating due to the vector meson decaying at rest. $l$ is the angular
momentum associated with twobody decay, and it is equal to unity for the
quoted $\Gamma_{V \to h_1h_2}$.
$B^2_l (X)$
denotes Blatt-Weisskopf barrier penetration factor. $B^2_{l=1} (X)$ is
given by $B^2_{l=1} (X) = X^2/(1+X^2)$ \cite{manl, pary0}.
The
dielectron decay width $\Gamma_{V \to e^+e^-} (m)$ is expressed in
Ref.~\cite{sibir}:
\begin{equation}
\Gamma_{V \to e^+e^-} (m)
\simeq \Gamma_{V \to e^+e^-} (m_V) \frac{m}{m_V}.
\label{wdee}
\end{equation}
The values of $\Gamma_{\omega \to e^+e^-} (m_\omega)$ and
$\Gamma_{\phi \to e^+e^-} (m_\phi)$ are 0.6 keV and 1.26 keV respectively
\cite{olive}.

\begin{figure}[h]
\begin{center}
\centerline {\vbox {
\psfig{figure=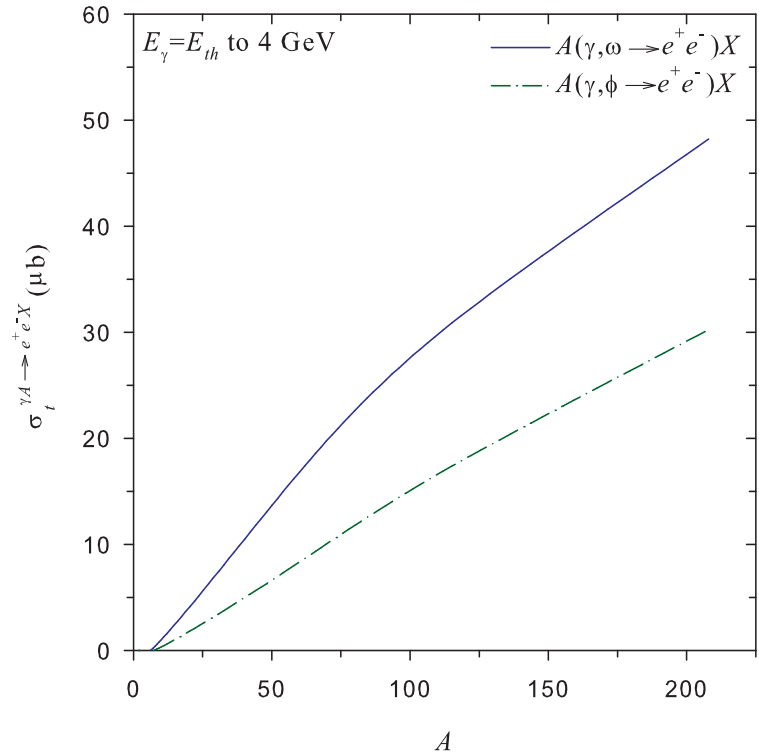,height=10.0 cm,width=10.0 cm}
}}
\caption{
(color online).
The calculated $A$ (mass number) dependent cross section of the
$\gamma A \to \omega X$; $\omega \to e^+e^-$ (solid line) and
$\gamma A \to \phi X$; $\phi \to e^+e^-$ (dot-dashed line) reactions.
}
\label{FgXT}
\end{center}
\end{figure}

The total cross sections of the $A(\gamma,V)X; ~V \to e^+e^-$ reaction have
been calculated to evaluate the nuclear transparency of the $\omega$ and
$\phi$ meson reported from Jlab \cite{wood}.
Therefore,
the range of vector meson mass in Eq.~(\ref{gAX2}) and the photon beam
energy range in Eq.~(\ref{gAX5}) are taken same as those mentioned in
Ref.~\cite{wood}.
The
calculated cross sections vs. $A$ (nuclear mass number) are presented in
Fig.~\ref{FgXT}, where the free space values of the $\omega$ meson nucleon
cross section $\sigma^{\omega N}_t$ and $\phi$ meson nucleon cross section
$\sigma^{\phi N}_t$ have been used.
The
solid curve denotes the total cross section of the above reaction because of
the $\omega$ meson, where as that due to the $\phi$ meson is illustrated by
the dot-dashed curve.
This
figure shows that the cross section of the previous reaction is larger
than that of the latter, and the cross sections of both reactions increase
with the size of the nucleus.

\begin{figure}[h]
\begin{center}
\centerline {\vbox {
\psfig{figure=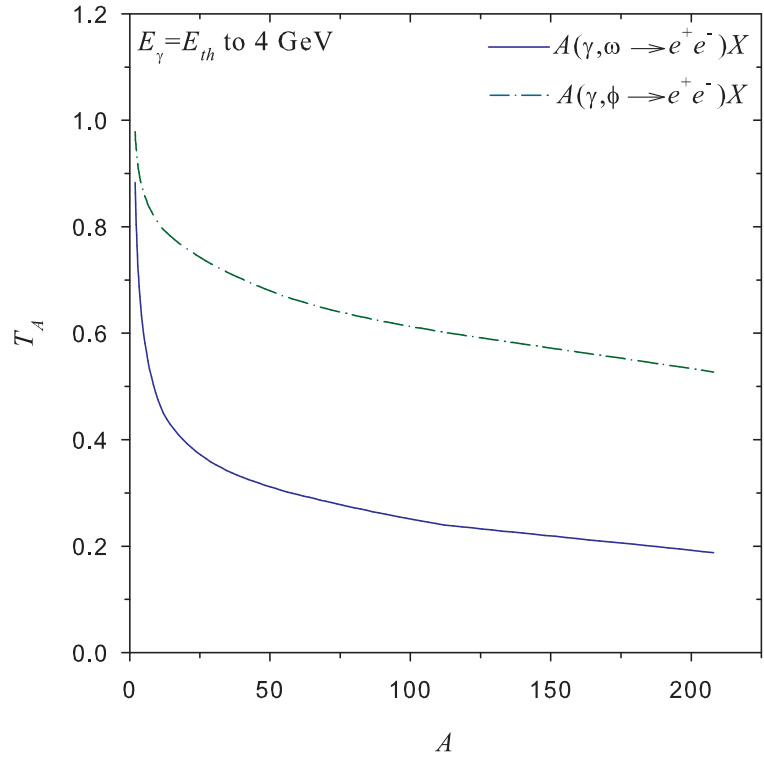,height=10.0 cm,width=10.0 cm}
}}
\caption{
(color online).
The calculated nuclear transparencies $T_A$ for the $\omega$ and $\phi$
mesons vs. mass number $A$ of the nucleus. The solid curve denotes $T_A$
for the $\omega$ meson, whereas the dot-dashed curve represents that for
the $\phi$ meson.
}
\label{FgTA}
\end{center}
\end{figure}

The nuclear transparencies $T_A$ of $\omega$ and $\phi$ mesons in the
above reaction are evaluated using Eq.~(\ref{ta}), and those have been
shown in Fig.~\ref{FgTA}.
The
solid and dot-dashed curves illustrate $T_A$ for the $\omega$ and $\phi$
mesons respectively. This figure shows larger absorption (lesser $T_A$) for
the $\omega$ meson compared to $\phi$ meson.
It
arises since the $\omega$ meson nucleon cross section $\sigma^{\omega N}_t$
(as shown in Table~1) is much larger than the $\phi$ meson nucleon cross
section $\sigma^{\phi N}_t$.

The calculated transparency ratio normalized to $^{12}$C nucleus, i.e.,
$T_A/T_C$, in the $\gamma A \to \omega X$; $\omega \to e^+e^-$ reaction are
compared with the data \cite{wood} in Fig.~\ref{FTOC}. The solid curve
describing $T_A/T_C$ for the $\omega$ meson, evaluated using
$\sigma^{\omega N}_t$, overestimates the data significantly.
Therefore,
the $\omega$ meson nucleon cross section in the nucleus
$\sigma^{*\omega N}_t$ is believed to be larger than $\sigma^{\omega N}_t$,
and the ratios $T_A/T_C$ are evaluated using different values of
$\sigma^{*\omega N}_t$ in Eq.~(\ref{ta}) instead of $\sigma^{\omega N}_t$.
The
calculated results for the latter are also presented in Fig.~\ref{FTOC}.
The short-dashed curve in this figure distinctly shows that $T_A/T_C$
calculated using the drastically enhanced
$\sigma^{*\omega N}_t (= 10\sigma^{\omega N}_t)$ could not reproduce the
data.

It is remarkable that Fig.~\ref{FTOC} shows a saturation of the transparency
ratio $T_A/T_C$ of $\omega$ meson for $\sigma^{*\omega N}_t \ge 5
\sigma^{\omega N}_t$. It originates because the $\omega$ meson of mean free
path in the nucleus $\lambda^*_\omega = 1/ (\varrho \sigma^{*\omega N}_t)$
considerably less than the nuclear radius $R$ cannot get out of the nucleus
except that produced at the surface of nucleus.
Once
the nucleus is black for the $\omega$ meson, it cannot be more black because
of the further increase in $\sigma^{*\omega N}_t$. This corresponds to the
limit: $\lambda^*_\omega /R << 1$. In that case, as shown by
M$\ddot{\mbox{u}}$lich and Mosel \cite{mueh}, the transparency $T_A$ reaches
the lowest possible value $\frac{\pi R^2}{A\sigma^{\omega N}_t}$, i.e., it
cannot fall faster than $A^{-1/3}$.

The dot-dashed curve in Fig.~\ref{FTPC} shows that the transparency ratio
$T_A/T_C$ for the $\phi$ meson calculated using $\sigma^{\phi N}_t$
overestimates the data \cite{wood} for the heavy nucleus.
This
figure also shows $T_A/T_C$, evaluated using the in-medium $\phi$ meson
nucleon cross section $\sigma^{*\phi N}_t = (1.5-10)\sigma^{\phi N}_t$,
yields very good agreement with the data for all nuclei.
To
be added, $T_A/T_C$ due to $\sigma^{*\phi N}_t = 3\sigma^{\phi N}_t$ (long
dashed curve) is consistent with $\sigma^{*\phi N}_t$ reported by Ishikawa
et al. \cite{ishi}, i.e., $\sigma^{*\phi N}_t \sim 35^{+17}_{-11}$ mb.

\section{Conclusions}

The cross section of the inclusive $ (\gamma,V \to e^+e^- )$ reaction on
nuclei has been calculated to evaluate the nuclear transparency $T_A$ of the
vector meson, i.e., $V \equiv \omega ~\mbox{or} ~\phi$ meson, in the
multi-GeV region.
The
interaction of the vector meson with the nucleus (optical potential) is
described by the elementary (free space) vector meson nucleon cross section
$\sigma_t^{V N}$.
The
nuclear transparency ratios normalized to $^{12}$C nucleus $T_A/T_C$
evaluated using $\sigma_t^{V N}$ overestimate the data (reported from Jlab)
for both $\omega$ and $\phi$ mesons.
This
ratio for the $\omega$ meson calculated using drastically enhanced $\omega$
meson nucleon cross section in the nucleus (i.e.,
$\sigma_t^{*\omega N} = 10\sigma_t^{\omega N}$) could not reproduce the data,
whereas $T_A/T_C$ for the $\phi$ meson evaluated using
$\sigma_t^{*\phi N} = (1.5-10)\sigma_t^{\phi N}$ reproduces the data.




\newpage

\begin{figure}[h]
\begin{center}
\centerline {\vbox {
\psfig{figure=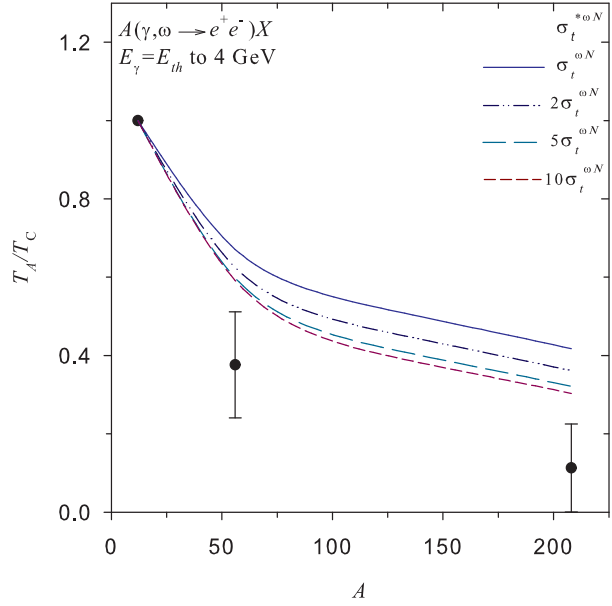,height=08.0 cm,width=08.0 cm}
}}
\caption{
(color online).
The nuclear transparency ratios (normalized to $^{12}$C) vs. $A$ (mass
number) for the $\omega$ meson calculated using the increasing values of
$\omega$ meson nucleon cross section in the nucleus.
$\sigma^{*\omega N}_t$
denotes the elementary $\omega$ meson nucleon total scattering cross section
in the nucleus, and $\sigma^{\omega N}_t$ represents that in the free space.
The data are taken from Ref.~\cite{wood}.
}
\label{FTOC}
\end{center}
\end{figure}

\begin{figure}[h]
\begin{center}
\centerline {\vbox {
\psfig{figure=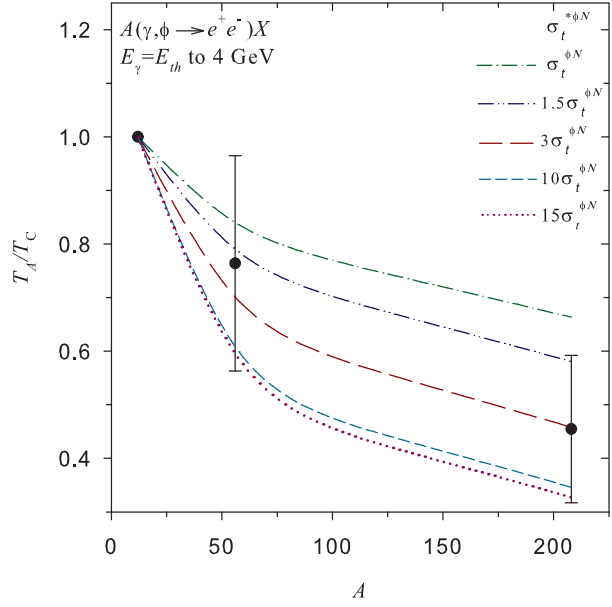,height=08.0 cm,width=08.0 cm}
}}
\caption{
(color online).
Same as those described in Fig.~\ref{FTOC} but for the $\phi$ meson.
$\sigma^{*\phi N}_t$ denotes the elementary $\phi$ meson nucleon total
scattering cross section in the nucleus, and $\sigma^{\phi N}_t$ represents
that in the free space. The data are taken from Ref.~\cite{wood}.
}
\label{FTPC}
\end{center}
\end{figure}

{\bf Figure Captions}
\begin{enumerate}

\item
(color online).
The calculated $A$ (mass number) dependent cross section of the
$\gamma A \to \omega X$; $\omega \to e^+e^-$ (solid line) and
$\gamma A \to \phi X$; $\phi \to e^+e^-$ (dot-dashed line) reactions.

\item
(color online).
The calculated nuclear transparencies $T_A$ for the $\omega$ and $\phi$
mesons vs. mass number $A$ of the nucleus. The solid curve denotes $T_A$
for the $\omega$ meson, whereas the dot-dashed curve represents that for
the $\phi$ meson.

\item
(color online).
The nuclear transparency ratios (normalized to $^{12}$C) vs. $A$ (mass
number) for the $\omega$ meson calculated using the increasing values of
$\omega$ meson nucleon cross section in the nucleus.
$\sigma^{*\omega N}_t$
denotes the elementary $\omega$ meson nucleon total scattering cross section
in the nucleus, and $\sigma^{\omega N}_t$ represents that in the free space.
The data are taken from Ref.~\cite{wood}.

\item
(color online).
Same as those described in Fig.~\ref{FTOC} but for the $\phi$ meson.
$\sigma^{*\phi N}_t$ denotes the elementary $\phi$ meson nucleon total
scattering cross section in the nucleus, and $\sigma^{\phi N}_t$ represents
that in the free space. The data are taken from Ref.~\cite{wood}.

\end{enumerate}

\newpage

\newpage

\newpage

\newpage

\end{document}